\documentclass[12pt,twocolumn]{iopart}
\usepackage{graphicx}
\usepackage{epsfig}
\usepackage{textcomp}
\usepackage{amsfonts}
\usepackage{amssymb}
\newcommand{\ket}[1]{|#1\rangle}

\newcommand{\unit}[1]{~\textrm{#1}}
\newcommand{\unitmicro}[1]{~\mu\textrm{#1}}
\begin{document}

\title[Coherent Manipulation]{Coherent Manipulation of a $^{40}$Ca$^+$ Spin Qubit in a Micro Ion Trap}
\author{U. G. Poschinger, G. Huber, F. Ziesel, M. Dei{\ss}, M. Hettrich, S. A. Schulz, K. Singer and F. Schmidt-Kaler
}
\address{Institut f\"{u}r Quanteninformationsverarbeitung, Universit\"{a}t Ulm,
Albert-Einstein-Allee 11, 89069 Ulm, Germany}

\ead{ulrich.poschinger@uni-ulm.de}

\author{G. Poulsen, M. Drewsen}
\address{QUANTOP - Danish National Research Foundation Center for Quantum Optics,
Department of Physics and Astronomy, Aarhus University, Denmark}

\author{R. J. Hendricks}
\address{Centre for Cold Matter, Blackett Laboratory, Imperial College London,
Prince Consort Road, London SW7 2AZ, United Kingdom}

\begin{abstract}
We demonstrate the implementation of a spin qubit with a single
$^{40}$Ca$^+$ ion in a micro ion trap. The qubit is encoded in the
Zeeman ground state levels $m_J=+1/2$ and $m_J=-1/2$ of the
S$_{1/2}$ state of the ion. We show sideband cooling close to the
vibrational ground state and demonstrate the initialization and
readout of the qubit levels with 99.5$\%$ efficiency. We employ a
Raman transition close to the S$_{1/2}$ - P$_{1/2}$ resonance for
coherent manipulation of the qubit. We observe single qubit
rotations with 96$\%$ fidelity and gate times below 5$\mu$s. Rabi
oscillations on the blue motional sideband are used to extract the
phonon number distribution. The dynamics of this distribution is
analyzed to deduce the trap-induced heating rate of 0.3(1)
phonons$/$ms.
\end{abstract}


%  37.10.Ty   Ion trapping
%  37.10.-x   Atom, molecule, and ion cooling methods
%  03.67.Lx   Quantum computation
%  32.80.Qk   Coherent control of atomic interactions with photons

\maketitle

\section{Introduction}

Our research is aimed at the realization of scalable quantum simulation  and
information processing ~\cite{CHUANG,ROADMAPS}. Quantum computing with cold
ions~\cite{ROOS1999,ROOSPHD,SCHMIDT2003,SEIDELIN2006} has currently reached  an
experimental limit of scalability with up to eight ions if a conventional macroscopic
trap is used~\cite{HAEFFNER2005}. This purely technical limitation is believed to be
lifted using a segmented linear micro Paul trap, where only small groups of ions are to
be kept in the quantum processing unit and multiple trap sites are used for the
storage of large-scale entanglement ~\cite{KIELPINSKI2002}. Several different options for
encoding a qubit with a trapped ion are possible and have been realized in various groups.
One might either employ superpositions of a long lived electronic metastable state and the ground state,
or alternatively use coherent superpositions of hyperfine or Zeeman ground states.
In this paper we focus on $^{40}$Ca$^+$ ions in a micro trap and qubits which are
encoded in Zeeman sublevels of the $S_{1/2}$ ground state. We will discuss in detail
how these qubits are initialized, coherently manipulated and how finally the quantum
information is read out with high fidelity. With the ion cooled close to its vibrational ground
state we are well set for two-qubit quantum logic gate operations in a
multi-segmented linear micro trap; a scalable approach to ion quantum processing.

The paper is organized as follows: First, we give a brief overview on the experimental
apparatus used including the micro trap and the various laser sources. In
Sect.~\ref{qubitvariations} we discuss the advantages of our choice of Zeeman ground state
superpositions for the qubit encoding. Two basic steps for qubit utilization, namely state preparation and
read-out are demonstrated in Sect.~\ref{qubitprepredout}. Finally, we explain in detail
how Raman transitions for the spin qubit manipulations are characterized
(Sect.~\ref{secraman}) and present results for sideband cooling and coherent qubit
dynamics (Sect.~\ref{sidebandcooling}). We show how phonon number distributions can be
extracted from the analysis of Rabi oscillations on the blue sideband of the Raman
transition and investigate the dynamics of the phonon distribution in the micro trap.
The outlook sketches the future perspectives of our experiment for multi qubit quantum
logic.

\begin{figure}[htp]\begin{center}
\includegraphics[width=0.8\textwidth]{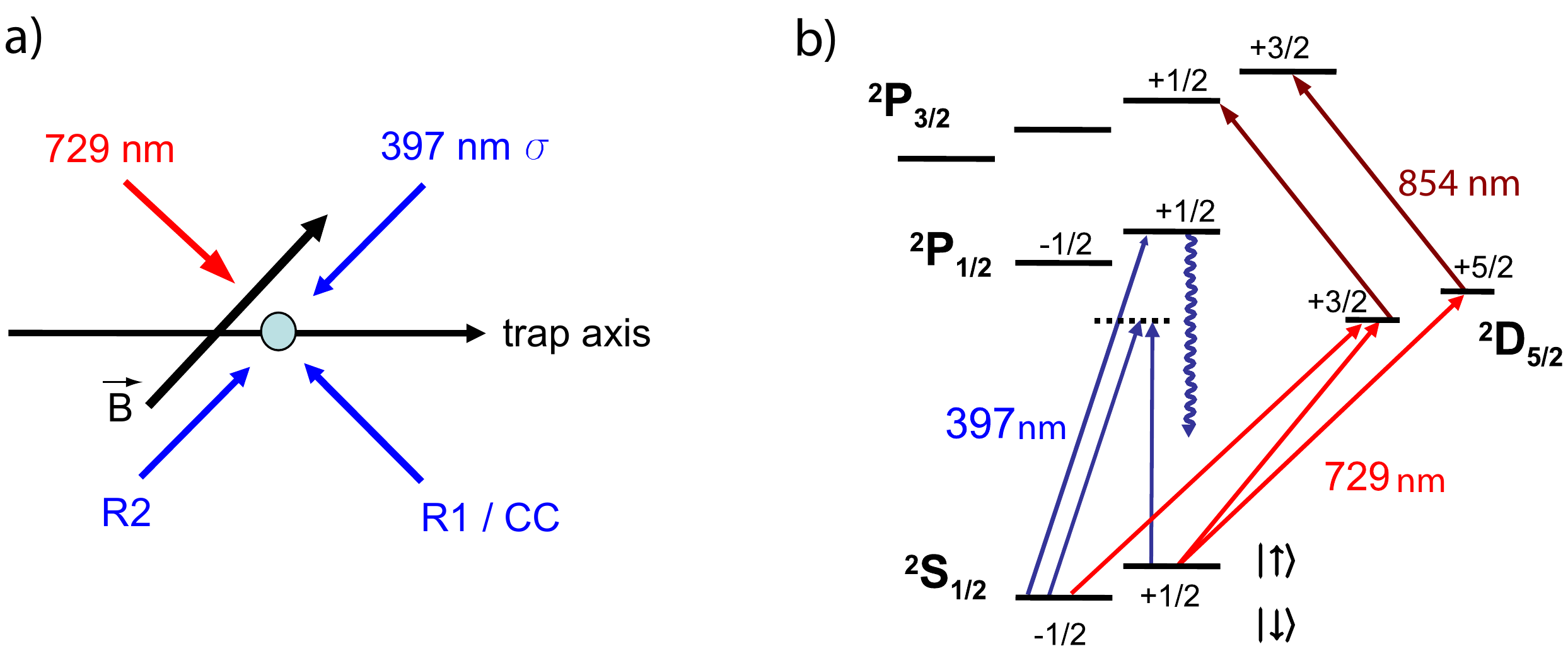}
\caption{ \textbf{a)} Geometric arrangement of the lasers driving Raman (R1, R2 and
CC), dipole ($397\unit{nm}$) and quadrupole transitions ($729\unit{nm}$), relative to the
trap axis and magnetic field direction $\vec{B}$. \textbf{b)} Level scheme of a
$^{40}$Ca$^+$ ion and all relevant transitions. A magnetic field of about $6.8\unit{G}$
 splits each fine structure level into several Zeeman components, indicated by fractional numbers.}
\label{levelscheme}
\end{center}\end{figure}

\section{Experimental Apparatus} \label{expapp}
\subsection{Segmented Micro Ion Trap}
We use a segmented linear micro ion trap. It is a sandwich structure of three alumina
wafers, of which the top and bottom ones are gold coated and the middle one acts as a
spacer. The trap structure is created by laser cutting with $\mu$m-scale resolution.
Each trap layer consists of a RF electrode providing ponderomotive confinement in
the radial plane and a set of DC electrodes for confinement in the axial direction.
The RF electrodes extend along the whole length of the trap and have notches at the positions
of the gaps between neighboring DC segments in order to suppress axial bumps in the RF
field. They are supplied by up to $300~\textrm{V}_\textrm{pp}$ at $24.8\unit{MHz}$. The
DC electrodes are supplied with voltages in $\pm10\unit{V}$ range by a
computer-controlled battery-powered supply which is designed to provide fast and
accurate voltage waveforms to the electrodes at low noise and output impedance. The
ionic motion exhibits one vibrational mode along the trap axis which is to be used as
the 'bus' mode for entangling operations. Under typical trapping conditions, when a voltage of -6 V is applied to one pair of dc segments and all other dc electrodes are grounded, an
axial  frequency of $\omega_{\rm ax}/2\pi = 1.35\unit{MHz}$ is measured. The radial
confinement leads to two nondegenerate radial modes with typical frequencies of
$\omega_{\rm rad}^{x,y}/2\pi = \{2.0,3.5\}\unit{MHz}$. For further details we refer to
the description in~\cite{SCHULZ2006,SCHULZ2008}.

\subsection{Laser Sources and Ion Detection}
A single ion is loaded by photoionization of a weak neutral calcium beam by resonant
two-photon photoionization~\cite{GULDE2001}. Doppler cooling is achieved by means of a
grating-stabilized diode laser at $397\unit{nm}$, slightly red detuned with respect to
the $S_{1/2}$ to $P_{1/2}$ dipole transition, see Fig.\ref{levelscheme}(a). As the
cooling cycle is not closed, a grating-stabilized diode laser at $866\unit{nm}$ is used
for repumping from the metastable $D_{3/2}$ state. The light from both sources is
switched by means of acousto-optical modulators (AOM). For Doppler cooling
we irradiate 100$\mu$W and 500$\mu$W near 397~nm and 866~nm, respectively. The waist size
of both beams is about 30$\mu$m at the ion's location. When necessary, population in the  $D_{5/2}$ state ($1.2\unit{s}$ lifetime) can be pumped to the $P_{3/2}$ by a grating stabilized laser diode
near 854~nm, followed by a fast decay to the S$_{1/2}$ ground state, which is referred to as quenching.
Since the long lifetime of the $D_{5/2}$ state is important for qubit measurement, no quench light must be present during readout.  To ensure sufficient extinction of the quench laser, the light of about $10\unitmicro{W}$ in a beam
waist of about $50\unitmicro{m}$ is switched by a double-pass AOM. All diode lasers
are locked to Pound-Drever-Hall (PDH) errors signals from stable Fabry-Pérot cavities
which are tuned by a piezo driven mirror.

In order to drive the $S_{1/2}$ to $D_{5/2}$ quadrupole transition we use an amplified diode
laser system.  This laser source near 729~nm is PDH locked to an ultra-low expansion
cavity in a UHV-vessel. We estimate the laser linewidth to be better than 5~kHz as
determined from Ramsey contrast measurements on the quadrupole transition, where
coherence decay times of up to $200\unitmicro{s}$ were observed. The laser light near
729~nm is switched and modulated by a double-pass AOM controlled by a versatile
function generator\footnote{{\it VFG 150}, Toptica Photonics} (VFG), a DDS/FPGA based
RF-source~\cite{WUNDERLICH2007}. This allows for generation of laser pulses with
almost arbitrary frequency, phase and amplitude profiles. The resulting laser beam has
a power of up to $140\unit{mW}$ and is focused down to roughly $10\unitmicro{m}$,
allowing Rabi frequencies up to $1.5\unit{MHz}$ on the $S_{1/2}$, $m_J=+1/2$ to
$D_{5/2}$, $m_J=+5/2$ transition. The polarization is chosen to be at an angle of
45$^\circ$ to the quantizing magnetic field, whereas the propagation direction is
orthogonal to it. This way, all allowed transitions between the Zeemann sublevels of
the $S_{1/2}$ and $D_{5/2}$ states become accessible~\cite{ROOSPHD}
(Fig.~\ref{levelscheme}). The beam enters at an angle of 45$^\circ$ with respect to
the trap axis, allowing for momentum transfer to the axial and the radial modes. We
calculate the coupling to the ion's motion (Lamb-Dicke factors) of $\eta_{\rm ax} =
0.06$ and $\eta_{\rm rad}^{x,y}=\{0.034,0.026\}$.

Raman transitions between the spin levels m$_J$ = +1/2 and -1/2 of the $S_{1/2}$ ground
state are driven by two laser beams close to the strong $S_{1/2}$ to $P_{1/2}$ dipole
transition. The beams are derived from a frequency doubled amplified laser diode system
delivering up to $120\unit{mW}$ of power at $397\unit{nm}$. The laser is divided into
three different beams which are termed R1, R2 and CC. The Raman transitions are then
driven either by the pair R1/R2 or the pair R1/CC, see Fig.~\ref{levelscheme}(b). All
Raman beams are switched and modulated by single-pass AOMs. The AOMs for R2 and CC are
supplied by the same VFG source as the AOM for the $729\unit{nm}$ laser. The AOM for R1
is supplied by an RF synthesizer, which also serves as a phase reference for the VFG,
therefore the necessary phase stability between R1/CC and R1/R2 is guaranteed. The
limit of accuracy for the relative frequency between VFG and synthesizer was measured
to be $10\unit{mHz}$, which has no adverse effect, because the timescale on which a
single measurement cycle is carried out is much shorter (up to $20\unit{ms}$).

The 397~nm fluorescence light emitted by the ion during Doppler cooling is detected by an electron multiplier CCD camera and a photomultiplier tube. With typical fluorescence rates of 50 counts
in 3~ms from a single ion with a background of about 5 counts from scattered light, we
can discriminate the state with a statistical error in the sub-per mil
range~\cite{ROOSPHD}.

All laser sources are controlled by a versatile
computer control program and continuously monitored by a wavemeter with
5~MHz accuracy. A typical experimental sequence consists of four steps: (a) The ion is
Doppler cooled for 3~ms, (b) then it is cooled close to the ground state of the axial
vibrational mode by resolved sideband cooling (Sect.~\ref{sidebandcooling}). (c) The
qubit is then initialized (Sect.~\ref{subsecpumping}) and (d) coherent manipulations are
performed on the Raman transition (Sect.~\ref{sec:Raman1}). Finally, (e) the population
in the m$_J$ = +1/2 qubit level is shelved (Sect.~\ref{subsecreadout}) to the D$_{5/2}$
level and (f) the state is read out by counting 397~nm laser induced fluorescence for
3~ms. After quenching the state by light near 854~nm the cycle (a) to (f) is repeated,
typically for 100 times, giving the average transfer probability on the qubit
transition.

\section{Qubit Realizations with Ca$^+$ Ions and Arguments for a Ground State Spin Qubit}
\label{qubitvariations} The level scheme of Ca$^+$ ions allows for at least three
options to encode qubit information in a long lived superposition of two electronic
quantum states. The \emph{optical qubit} is encoded in a superposition of the ground
state $|0\rangle \equiv$ S$_{1/2}$ and the metastable $|1\rangle \equiv$ D$_{5/2}$.
Coherent manipulations are driven directly on this quadrupole transition by laser pulses
near 729~nm, an approach has been realized with great success by the Innsbruck group
\cite{MONZ2009,KIRCHMAIR2009}. Disadvantages to this approach are limitations to the coherence due to phase
stability of the laser source at 729~nm and ambient magnetic field fluctuations, as well
as the relatively small Lamb-Dicke factor which leads to a rather weak momentum kick of
the laser excitation on the ion vibrational motion, affecting the two qubit gate speed.
The optical qubit is read out efficiently and simply by state dependent fluorescence, as
the ground state scatters photons while the D$_{5/2}$ remains dark.

Another option is to use $^{43}$Ca$^+$ with nuclear spin I=7/2 and to encode the qubit in
\emph{hyperfine ground state levels} $|F=4,m_F=0\rangle$ and $|F=3,m_F=0\rangle$. Coherent
manipulations are achieved by employing a Raman transition, which means that an ultra-stable laser source is no longer required.  It is straightforward to achieve excellent relative phase stability if the two Raman beams are derived from a single laser source.  However, the large frequency gap of about 3.4~GHz must be bridged with high-bandwidth AOMs. Hyperfine clock states, insensitive to the linear Zeeman effect, can be used as computational basis states \cite{BENHELM2008}, which greatly
increases the qubit phase coherence. The qubit readout is based on state dependent
fluorescence.

For the measurements presented here we have chosen the option to encode the \emph{qubit
in spin sublevels} $\ket{\!\downarrow }$ and $\ket{\!\uparrow }$ of the ion's electronic
ground state S$_{1/2}$ m$_J$ = $\pm$ 1/2 of $^{40}$Ca$^+$ with I=0, Zeeman split by an
applied magnetic field. Coherent manipulations are achieved by employing a Raman
transition and the requirement of an ultra-stable laser source can then be dropped as
in the case of hyperfine qubits, but the much smaller frequency splitting of about 20~MHz allows the use of simple and efficient AOMs. The high Lamb-Dicke factor of UV-Raman transitions ensures
fast two qubit gate operations. As compared to ion species with hyperfine structure we
work with a much simpler level structure reducing the experimental complications. Qubit readout, however, is more complicated since both the $\ket{\!\downarrow }$ and $\ket{\!\uparrow }$ states can scatter photons near 397~nm.  In order to discriminate between the two qubit states, the population in $\ket{\!\uparrow }$  must first be completely transferred to the metastable D$_{5/2}$ state. A
future possibility to avoid qubit dephasing due to its magnetic field sensitivity for
the $^{40}$Ca$^+$ spins will be using two physical qubits (ions), in a decoherence
free subspace of Bell states to encode one logical qubit \cite{DFSQUBIT,HAEFFNER20052}, in the spirit of
designer atoms~\cite{ROOS2006}.

\section{Qubit Preparation and Readout} \label{qubitprepredout}
According to the DiVincenzo criteria~\cite{DIVICENZO2000}, the initialization of qubits
to a well defined state and the read-out of the full quantum state of the qubits are
essential criteria for the realization of quantum information experiments. In our
experiment, both steps are realized by using a narrow optical quadrupole transition at
$729\unit{nm}$, and both are optimized to achieve high fidelities even under the
presence of experimental imperfections and noise.

\subsection{State Preparation via Optical Pumping} \label{optpump}
\label{subsecpumping} The common technique for preparing an initial quantum state via
optical pumping employs a circularly polarized laser beam resonant with the S$_{1/2}$ to
P$_{1/2}$ dipole transition at $397\unit{nm}$. In this case, both double-refraction by
the vacuum windows and a small offset angle between the propagation direction
$\vec{k}_{397}$ and the quantizing magnetic field results in spurious polarization
components. The improper polarization components deteriorate the initialization
fidelity. Furthermore, as the target state is not completely dark anymore, this leads to a continuous
photon scattering of Doppler cooling light, which counteracts the ground state sideband
cooling (Sect.~\ref{sidebandcooling}). Therefore, we use the high spectral selectivity of
the narrow quadrupole transition for optical pumping \cite{ROOS2006}. If the ion is to
be initialized in the $\ket{\uparrow}$ level, the population from $\ket{\downarrow}$ is
transferred to the  $D_{5/2}$, $m_J=+3/2$ level by a pulse at $729\unit{nm}$ and
transferred back to the ground state by the quench laser via the $P_{3/2}$ state. This
cycle is repeated until the desired initialization is reached with high fidelity.

We compare two different schemes for this pumping: Either we use a pulsed scheme or we
switch on both light fields continuously on (Fig.~\ref{pumping}). In the pulsed scheme,
after about a $\pi$ pulse on the quadrupole transition the quench beam at 854nm is briefly switched on.  We find that a quench pulse of just 2$\unitmicro{s}$ is sufficient for complete
quenching of the D$_{5/2}$ state.

The $729\unit{nm}$ $\pi$-pulse length of about $10\unitmicro{s}$ determines the amount of
off-resonant excitation and thus the fidelity of the scheme: The target transition
$\ket{\uparrow}$ to $D_{5/2}$, $m_J=+5/2$ is separated from the parasitic transition
$\ket{\downarrow}$ to $D_{5/2}$, $m_J=+3/2$ by $\sim 8\unit{MHz}$ for a magnetic field
of 6.7~G. The frequency component of the Fourier transform of the effective
square pulses (without any pulse shaping) already results in an expected pumping
fidelity better than 99.6\%, in agreement with the value for the combined pumping and
readout fidelity (see below) in the experiment of 99.6\% {Fig.~\ref{pumping}a). The
cw scheme suffers from the fact that the presence of the quench beam hinders the
coherent buildup of population in the $D_{5/2}$ state \footnote{The quench laser
couples the metastable state to the rapidly decaying $P_{3/2}$ state. Upon decay from
this state, a photon is emitted which can be in principle detected, indicating that the
system has ended up in the ground state. This represents an effective continuous
measurement, disturbing the coherent evolution of the $S_{1/2}$-$D_{5/2}$
superposition.}

\begin{figure}[htp]\begin{center}
\includegraphics[width=0.95\textwidth]{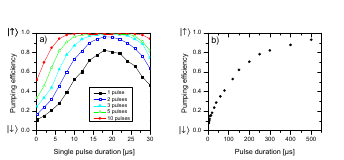}
\caption{ Single ion population is
pumped by the application of $\sigma$ - light at 397~nm to the $\ket{\!\downarrow }$
state. From this state we use optical pumping by laser light at 729~nm and 854~nm to
pump to the $\ket{\!\uparrow }$ via the $D_{5/2}$, $m_J=+3/2$ level. a) Pumping
efficiency of the the pulsed scheme versus the $729\unit{nm}$ pump time of the
individual pulses. The curves for various pulse numbers show how performance and robustness
increase for more pulses. Note that the total duration of the pump process increases with the number
of pulses. b) The cw-scheme is found less efficient.} \label{pumping}
\end{center}\end{figure}

\subsection{Spin readout} \label{subsecreadout}
A simple readout of the spin state by fluorescence observation is impossible
because the Zeeman splitting of the spin level $\ket{\!\uparrow }$ and $\ket{\downarrow
}$ is smaller than the natural linewidth $\Gamma/2\pi\approx 22\unit{MHz}$ of the
corresponding dipole 397~nm transition. A scheme circumventing this by means of electromagnetically induced transparency has been proposed and experimentally realized \cite{MCDONNELL2004}, reaching a fidelity of 86\%.

Our scheme reaches 99.6$\%$ fidelity and additionally shows a high robustness against
imperfect laser settings or noise in the control parameters, still with a remarkably
modest experimental effort. In a first step we apply a rapid adiabatic passage pulse
(RAP)~\cite{WUNDERLICH2007} where the amplitude is adiabatically switched on and off
and the frequency is chirped across resonance\footnote{As stability against Rabi
frequency errors is the main issue for readout robustification, we have also employed
the SCROFOLOUS technique~\cite{CUMMINS2003,TIMONEY2008} which is robustified by using a
series of three $\pi$-pulses with different phases. Although the technique was found to
yield the same performance and an enhanced robustness against pulse area errors, the
low resilience against the laser frequency drift strongly limited the practical use.}.
Even for an ion after Doppler cooling, a single RAP pulse on the $\ket{\uparrow}$ to
$D_{5/2}$, $m_J=+5/2$ transition yields a fidelity of 95\%, the leftover population in the
$\ket{\uparrow}$ is then transferred by a second RAP to the $D_{5/2}$, $m_J=+3/2$ state and we
reach a 99.6$\%$ readout fidelity with a high resilience against drift of 729~nm laser frequency,
see Fig.~\ref{readout}.

\begin{figure}[htp]\begin{center}
\includegraphics[width=0.7\textwidth]{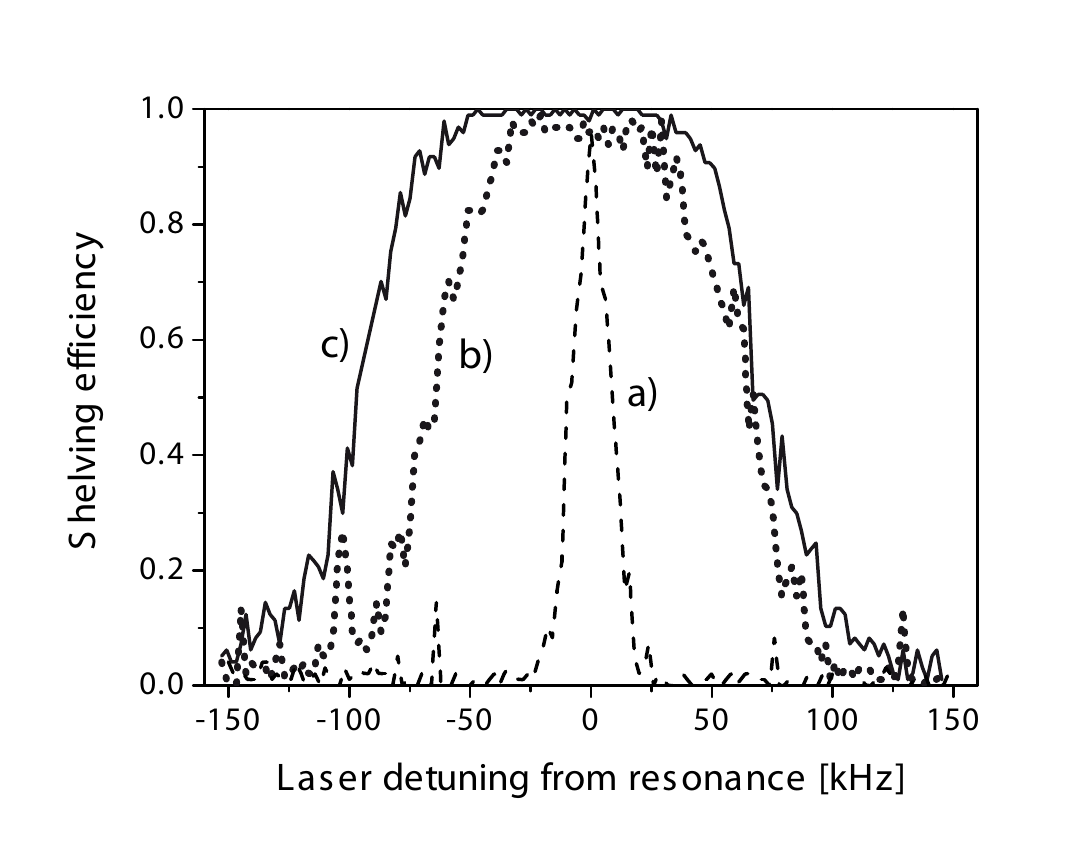}
\caption{Transfer efficiency of the spin qubit $\ket{\uparrow}$ for various schemes
versus central detuning of the readout pulse. a) Dashed - the narrow peak is the result
for a Gaussian pulse with a resonant pulse area of $\pi$. b) Dotted - the much broader peak corresponding to a single RAP pulse shows that the robustness against frequency
errors has greatly improved. c) Solid - the curve for the double RAP shows that both
performance and robustness have increased with respect to the single RAP.}
\label{readout}
\end{center}\end{figure}

\section{Raman Transitions between the Spin Qubit Levels} \label{secraman}

\subsection{Raman Spectroscopy and Characterization of the Transition} \label{sec:Raman1}
In our experiment we use Raman beams derived from one laser source (see
Sect.~\ref{expapp}). The laser is detuned from the $S_{1/2}\to P_{1/2}$ resonance
frequency by $\Delta$, referred to as the Raman detuning. It is up to several tens of
GHz and can be chosen to be both positive (blue detuning) or negative (red detuning).
As $\Delta$ is very large compared the Zeeman splittings within the $S_{1/2}$ and
$P_{1/2}$ manifolds, which is on the order of 10 to 20 MHz, it
can be considered constant for the different transitions between the Zeeman levels. The
$P_{1/2}$ state can be adiabatically eliminated from the dynamics, giving an effective
two-level system. The Raman Rabi frequency now reads
$\frac{\Omega_1\Omega_2}{2\Delta}$, where $\Omega_i$ are the resonant single dipole Rabi
frequencies associated with the individual beams.

A great advantage of the utilization of Raman transitions for quantum logic is the
better control over the Lamb-Dicke factor. The difference k-vector of the beams
strongly depends on the chosen geometry, see Fig.~\ref{levelscheme}(a). In our setup,
two different beam geometries are employed.  In the first of these, a pair of Raman beams, R1 and CC, propagate parallel to each other and orthogonal to the magnetic field.  The difference k-vector of the two beams is effectively zero, so the Lamb-Dicke factor is extremely small and electronic excitation is insensitive to the ion's motion.
One of the beams (referred to as 'R1') is $\pi$-polarized, driving the $\Delta m_J=0$ transition. The
other beam (referred to as 'CC', cocarrier) is polarized orthogonal to the magnetic field. It therefore
yields two circular components, one of which contributes to the coherent dynamics of
the effective two-level system if the Raman resonance condition is fulfilled, i.e. if
the relative detuning of the beams matches the Zeeman splitting between the qubit
levels. The alternative geometry consists of R1 and a beam propagating along the magnetic
field direction (referred to as 'R2') such that the k-vector of the beat pattern is aligned along the
trap axis. With no component of the k-vector in the radial plane of
the trap, this Raman light field interacts only with the axial vibrational mode, and
avoids any phonon induced dephasing from the radial modes commonly called spectator
modes. This fact is of tremendous importance for our microtrap with prospects to many
ions, as we can drop the requirement of ground state cooling of all vibrational modes which would
be experimentally undesirable.

\begin{figure}[htp]\begin{center}
\includegraphics[width=0.95\textwidth]{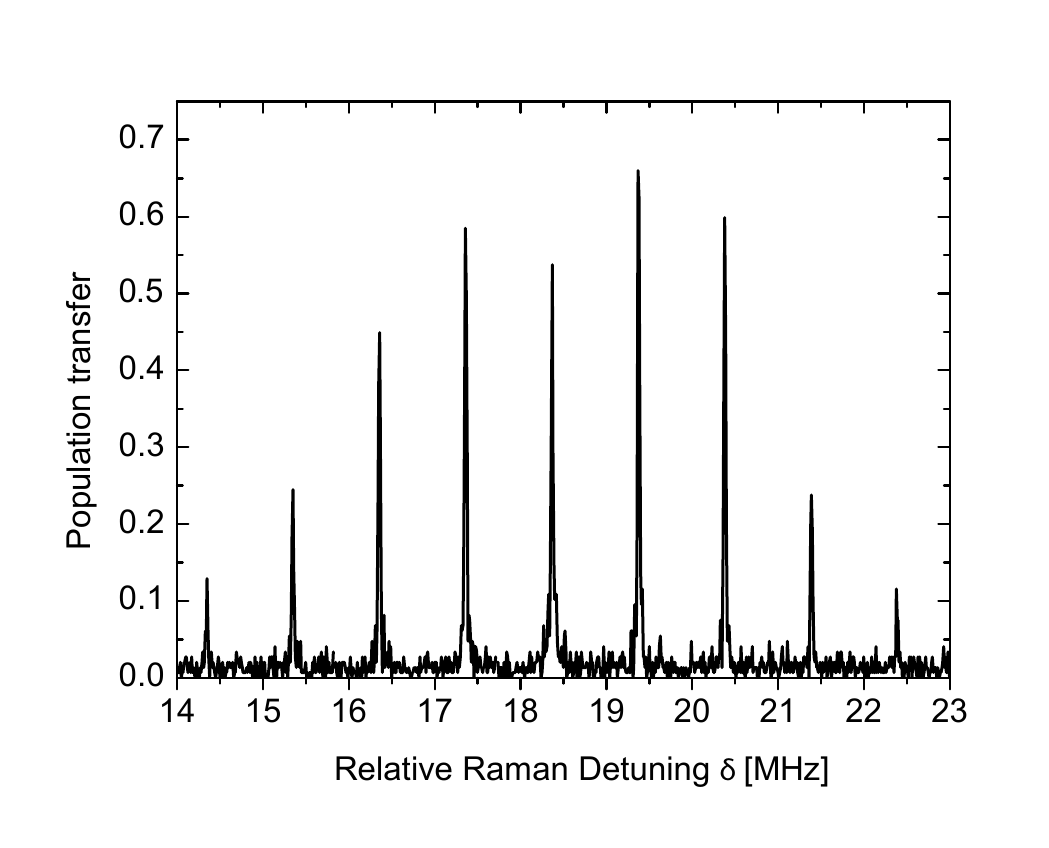}
\caption{Raman spectrum obtained as R1 is fixed in frequency and R2 is scanned over the
sideband resonances. The excitation geometry leads to a difference k-vector along the
trap axis, such that only axial, but no radial modes show up.} \label{spectrum}
\end{center}\end{figure}

From the beam geometry and the axial vibrational frequency we deduce a Lamb-Dicke
factor of $\eta = 0.21$, which is much larger than for the optical transition.

A Raman spectrum is shown in Fig.~\ref{spectrum}, clearly displaying the axial sidebands
of motion for a single ion after Doppler cooling with $\bar{n}\sim$ 15 quanta and an axial
trap frequency of 1.35~MHz.

\begin{figure}[htp]\begin{center}
\includegraphics[width=0.75\textwidth]{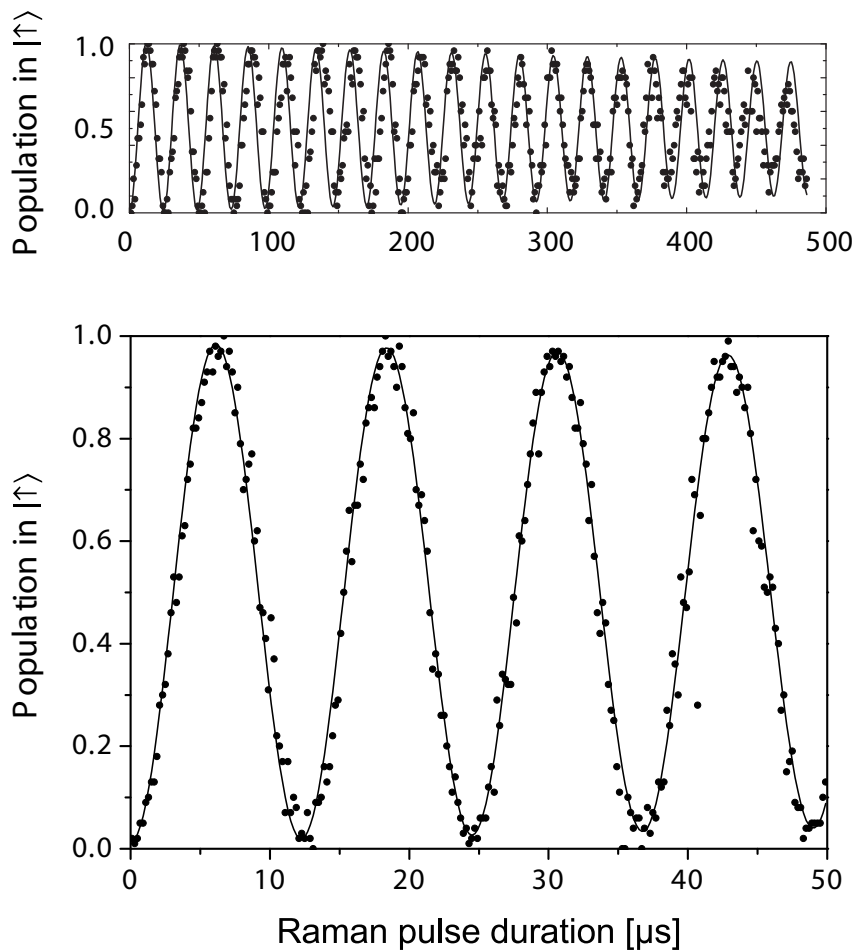}
\caption{Rabi oscillations on the Raman transition driven by the R1/CC pair. No phonon
induced dephasing occurs because the Lamb-Dicke factor is virtually zero, which allows
for driving many single qubit rotations with high fidelity. The contrast of the
oscillation is 96\%. The upper curve shows the long coherence time, while the lower
curve displays the first four Rabi oscillations with high resolution.} \label{ccrabi}
\end{center}\end{figure}

\subsection{Raman Rabi Oscillations on the Carrier Transition}
Rabi oscillations driven by the R1/CC beams pair are shown in Fig.~\ref{ccrabi}. Here,
with $\eta = 0$, no phonon-induced dephasing can occur. For the experiment, we follow
the sequence in Sect.~\ref{expapp} without ground state cooling in step (b) and apply
in step (c) both beams R1 and CC simultaneously for an interaction time $t$. The
experiment is repeated 100 times and the average excitation probability is plotted.

Showing no influence of the phonon number distribution, this technique provides an excellent opportunity for studying other sources of decoherence. These are ambient magnetic field fluctuations,
fluctuations of the relative phases of the beams due to air currents or mechanical
vibrations, laser intensity fluctuations and spontaneous photon scattering. The latter
two mechanisms scale with the total effective pulse area imparted to the ion on each of
the dipole transitions pertaining to the Raman transition. Because of this, a Ramsey
contrast measurement on a spin superposition created by R1/CC with fixed pulse areas
and variable delay between the Ramsey pulses allows us to study the limits imposed by magnetic field noise and interferometric stability.  Such a measurement has
given a contrast of 90\% after a delay time of 1.5~ms. As $\pi$-times of down to
$2\unitmicro{s}$ are routinely achieved, this separation of time scales appears
sufficient for quantum logic experiments. We find in our setup that an air shield and
the high passive mechanical interferometer stability of the optical setup are
sufficient to avoid technically induced dephasing. The fundamental physical coherence
limit of qubits is given by spontaneous photon scattering ~\cite{OZERI2005,OZERI2007}
\footnote{It should be mentioned that in contrast to the work in
\cite{OZERI2005,OZERI2007}, where hyperfine levels were used for qubit encoding, the
spin qubit yields the advantage that decoherence during off-resonant entangling gate
operations is greatly reduced. This is due to the fact that if a photon is scattered, it
does not carry any information about the final state out of the system as the frequency
splitting of the qubit levels is on the same order as the linewidth of the
corresponding dipole transition.}.

During operations with the Raman beams, the small amount of population off-resonantly excited to the $P_{1/2}$ state decays with a rate corresponding to the inverse lifetime of this state, which results in random
repopulation of the qubit levels. This off-resonant excitation reads for each beam $i$ as
$p_{P_{1/2}}=\Omega_i^2/2\Delta^2$, decreasing with the ratio of Rabi frequency and Raman
detuning. Similar to far detuned optical dipole traps \cite{RUDI}, the photon
scattering rate can be reduced if a large $\Delta$ is chosen at a high laser intensity.
We investigated the photon scattering experimentally: For a given Raman detuning and
fixed beam powers, the Rabi frequency is measured from Rabi oscillations as in
Fig.~\ref{ccrabi} along with the scattering rates caused by each of the Raman beams
individually. For this, the ion is initialized in $\ket{\uparrow}$ and one
of the Raman beams is blocked. We apply the remaining Raman beam with a pulse of
variable length. Finally, from the population in $\ket{\downarrow}$ we infer the scattering rate of beam $i$.

\begin{figure}[htp]\begin{center}
\includegraphics[width=0.6\textwidth]{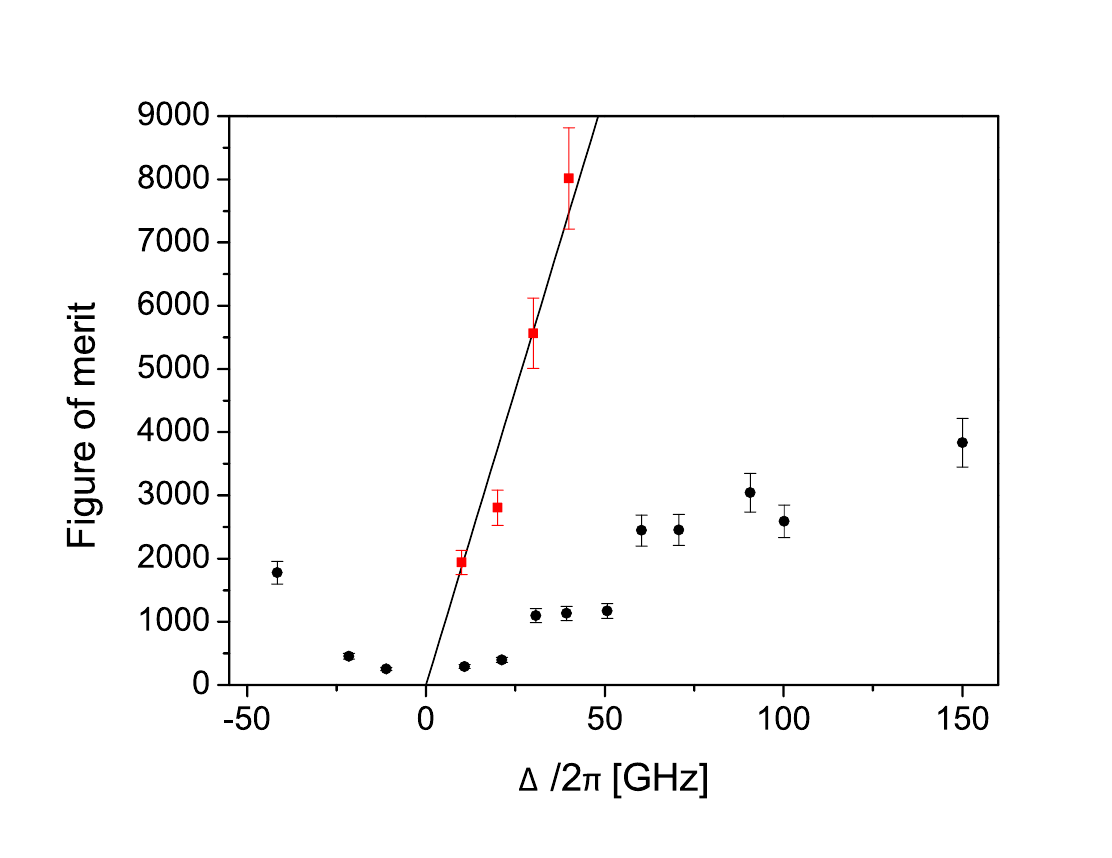}
\caption{Rabi frequency divided by geometric mean of the scattering rates versus Raman
detuning for the R1/CC pair (red squares) and the R1/R2 pair (black dots). The solid
line through the R1/CC data is the theoretical value of $\sqrt{18}\Gamma^{-1}|\Delta|$.}
\label{scatter}
\end{center}\end{figure}

For the analysis, we normalize the measured Rabi frequency by the geometric mean of
both scattering rates, such that in the resulting quantity, the individual Raman beam dipole coupling strenghts
$\Omega_i$ are canceled.

A value of $\sqrt{18}\Gamma^{-1}|\Delta|$ is expected when the Clebsch-Gordan coefficients
for the transitions are taken into account. The result
is shown in Fig.~\ref{scatter}, where the vertical axis can be interpreted as the number of Rabi
cycles that could be driven on average  before a single spontaneous scattering event occurs, in the absence of any other decoherence sources. The results for the R1/CC pair match very well the theoretical expectation. The fact that the values for the R1/CC pair match the theoretical expectation was initially not taken for
granted because of the type of laser source used: A tapered amplifier generates a
background of amplified spontaneous emission whose width in the range of a few nm is
given by the gain profile of the semiconductor laser medium. Through sum frequency
generation, it is in principle possible that photons at frequencies
$\omega_{0}+n\;\omega_{FSR}$ are generated in the doubling stage, where $\omega_{0}$ is
the laser mode frequency and $\omega_{FSR}$ is the free spectral range of the doubling
stage. If such a frequency matches the direct optical $S_{1/2}$ to $P_{1/2}$
transition, resonant photon scattering would occur.

The same experiment, but with the R1/R2 pair of Raman beams, yields lower values for
the Raman Rabi frequency. As in this geometry the excitation is sensitive to the
motional degree of freedom along the trap axis, we attribute the lower value to the
decrease of the Raman Rabi frequency with increasing phonon number and axial micromotion
components \footnote{The presence of axial micromotion was confirmed by checking for RF-echoes in the Raman
spectrum of the R1/R2 pair, i.e. Raman resonances at relative detunings of
$\omega_{Zeeman}+n~\omega_{RF}$. Note that the effect of micromotion on the Raman Rabi frequency has even been used to allow the fine tuning of Rabi frequencies in quantum gate experiments with Be$^+$ ions
\cite{TURCHETTE98}} at the trap drive frequency of $24.8\unit{MHz}$.

Another characterization of the Raman interaction is done by the determination of the AC-Stark
shifts, given by $\Delta_S=\Omega_i^2/4\Delta$ for one beam $i$. The absolute
AC Stark shift might be different for the two qubit levels $\ket{\uparrow}$ and
$\ket{\downarrow}$, leading to a differential shift. Under these circumstances, the
quantum phase of qubit superpositions, or of multi-ion entangled states evolves not
only according to the desired gate operations, but shows an additional intensity
dependent rapid phase evolution. If this is the case, intensity fluctuations of the Raman lasers
lead to qubit phase fluctuations, which represents an additional strong decoherence source.

The absolute AC-Stark shifts for all beams and qubit states is measured with the following procedure:
First, the qubit is prepared either in the $\ket{\!\uparrow }$ or the $\ket{\!\downarrow }$ level by optical pumping with circularly polarized 397~nm laser light. Then, a superposition on the quadrupole transition is created by
means of a $\pi/2$ pulse with the 729~nm laser, either with the $D_{5/2}, m_J=+5/2$
level for the $\ket{\!\uparrow }$ state or  $D_{5/2}, m_J=+3/2$ level for $\ket{\downarrow
}$. A second $\pi/2$ pulse after a delay time of $50\unitmicro{s}$ concludes the Ramsey
sequence. During this delay, a phase shift pulse from one of the Raman beams is irradiated on
the ion, leading to a Ramsey fringe signal as the duration of the shift pulse is
scanned. The absolute shift is then calculated from the fringe period $t_R$ according to
$\Delta_S=2\pi/t_R$. For the R1 beam the $\ket{\!\uparrow }$ level is shifted by
$2\pi\cdot 0.32(2)$MHz, and the $\ket{\!\downarrow }$ level is shifted by $2\pi\cdot
0.33(2)$MHz. Within the experimental errors, the differential shift from R1 vanishes.
In contrast, we measured shifts from the R2 beam with $2\pi\cdot 0.17(2)$MHz for the
$\ket{\!\uparrow }$ level and $2\pi \cdot 0.29(2)$MHz for the $\ket{\!\downarrow }$ level,
resulting in a differential shift of about $2\pi \cdot$120~kHz. Compensation of
the differential shift is possible by proper adjustment of the polarization of the corresponding
beam.
%For a compensation of the differential shift we balance the circular polarization
%components, e. g. if in the R2 beams one of the circular components is stronger than
%the other, the qubit level which is coupled to the $P_{1/2}$ manifold by this component
%experiences a stronger energy shift.

\subsection{Resolved Sideband Cooling and Blue Sideband Rabi Oscillations}
\label{sidebandcooling}

Cooling close to the ground state of at least one vibrational mode is an essential
prerequisite for two-ions gates, as even gate schemes for 'hot' ions require operation
in the Lamb-Dicke regime $\eta\;sqrt{n} \ll  1$~\cite{LEIBFRIED2002}.  For cooling close to the
ground state one has to resort to a narrow transition with resolved motional
sidebands~\cite{MARZOLI}, such that transitions to states with one less phonon (red
sideband, RSB) can be driven preferentially and the $n=0$ state acts as a dark
state in which the population is finally trapped. In our system, we have two options to
spectroscopically resolve sidebands, either the R1/R2 Raman transition or the
quadrupole transition.

As the cooling always competes with the heating rate from trap induced electric field
noise, a high cooling rate is essential for a good cooling result. A priori, the Raman
transition seems to be better suited for cooling because the higher ratio of RSB to
carrier Rabi frequency, which is essentially given by $\eta$ on the decisive
'bottleneck' step from $n=1$ to $n=0$. This is because ideally, the cooling rate is
limited by off-resonant excitation of the carrier transition with subsequent
spontaneous decay which can lead to the creation of one phonon. However, the problem arises in the
dissipative step of cooling where the ion is repumped to the initial state
to restart the red sideband excitation. In the case of the Raman cooling scheme, the
repumping is accomplished by the circular $397\unit{nm}$ beam which suffers from
the spurious polarization error discussed in Sect.~\ref{optpump}. Therefore the dark state
$n=0$ is not completely dark anymore, leading to a competing Doppler re-heating process
which limits the attainable temperature.

In contrast, the sideband cooling on the $\ket{\!\uparrow }$ to $D_{5/2}, m_J=+5/2$ quadrupole transition does not suffer from this because the repumping is achieved by the quench laser near 854~nm, which does not interact with
the ion anymore once one photon is spontaneously scattered. The cooling cycle is almost
closed, because the decay from the $P_{3/2}$ state during the quench leads
preferentially to the $\ket{\!\uparrow }$ level due to the selection rules. Only unlikely decay
events into one of the D-states can lead to population of the $\ket{\!\downarrow }$ level. We utilize a pulsed
sideband cooling scheme, since as for the qubit initialization, the power and frequency
of the quench laser are no longer critical parameters then. The cooling pulse time is set such
that an excitation maximum is reached on the RSB. This time ranges typically between
$10\unitmicro{s}$ and $20\unitmicro{s}$, and increases as lower phonon numbers are
reached because the RSB Rabi frequency scales as $\eta_{\rm ax}\sqrt{n}$ with the phonon
number $n$. After the RSB pulse, a quench pulse of typically $2\unitmicro{s}$ completes
the cooling cycle. After ten cooling cycles, about 10\% of the population is accumulated in
the wrong ground state spin level, such that a $397\unit{nm}$ repump pulse has to be
employed.  After eight such sequences, we employ a second cooling stage where the RSB
pulse duration is increased and the $729\unit{nm}$ optical pumping procedure is used
instead of the circular $397\unit{nm}$  pulses. The longer time for repumping has no
adverse effect on the cooling rate because it is used only every ten cycles.

\begin{figure}[htp]\begin{center}
\includegraphics[width=1.0\textwidth]{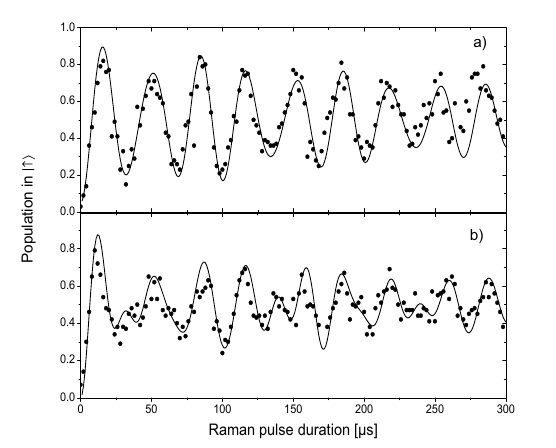}
\caption{Coherent dynamics on the R1/R2 BSB after sideband cooling. The graphs show the
population in the $\ket\downarrow$ level versus pulse duration of a square Raman
excitation pulse a) directly after sideband cooling and b) after a delay of 3~ms. The
data were obtained with a Raman detuning of $\Delta\approx 40~GHz$. The solid lines are
reconstructed from the extracted phonon distribution data with inclusion of a coherence
decay time of 280~$\unitmicro{s}$. We extract a mean phonon number of ~0.24 for
the data set without waiting time.} \label{coolstuff}
\end{center}\end{figure}

\subsection{Determination of the Phonon Number Distribution} \label{phononnumber}
We confirm the sideband cooling result by employing either the quadrupole
transition or the R1/R2 Raman transition. The optimization of the cooling is
performed by minimizing the peak excitation of the RSB of the quadrupole transition,
which is essentially given by the probability of not finding the ion in the ground state.
For more accurate determination of the phonon number distribution we employ Raman
Rabi oscillations on the R1/R2 BSB, with the advantage that no contributions from the
radial vibrational modes can influence the result, and on the other hand the larger
Lamb-Dicke factor of the Raman transition leads to a better separation of the Rabi
frequencies for the various $n\rightarrow n+1$ transitions. Excitation data are
acquired until the oscillation contrast of the excitation signal has decreased beyond
the projection noise limit for long pulse widths, see Fig.~\ref{coolstuff}. The recorded
traces are analyzed by cosine transform to obtain the frequency components for the
different contributing transitions, in full analogy to experiments on the cavity QED
realization of the Jaynes-Cummings model~\cite{BRUNE1996} \footnote{Due to the finite
data acquisition time, the peaks in the cosine transform pertaining to a given
transition frequency are accompanied by aliases at other frequencies which lead to
systematic errors when the phonon number occupation probability is inferred directly from the
peak heights. A deconvolution procedure was used to remove this effect. The correctness
of the method is proven by the fact that the method yields the correct input phonon number
distribution when Monte-Carlo generated data with realistic parameters is used. The resulting accuracy is then limited by the read-out projection noise of the pulse width scan data.}.

\begin{figure}[htp]\begin{center}
\includegraphics[width=0.9\textwidth]{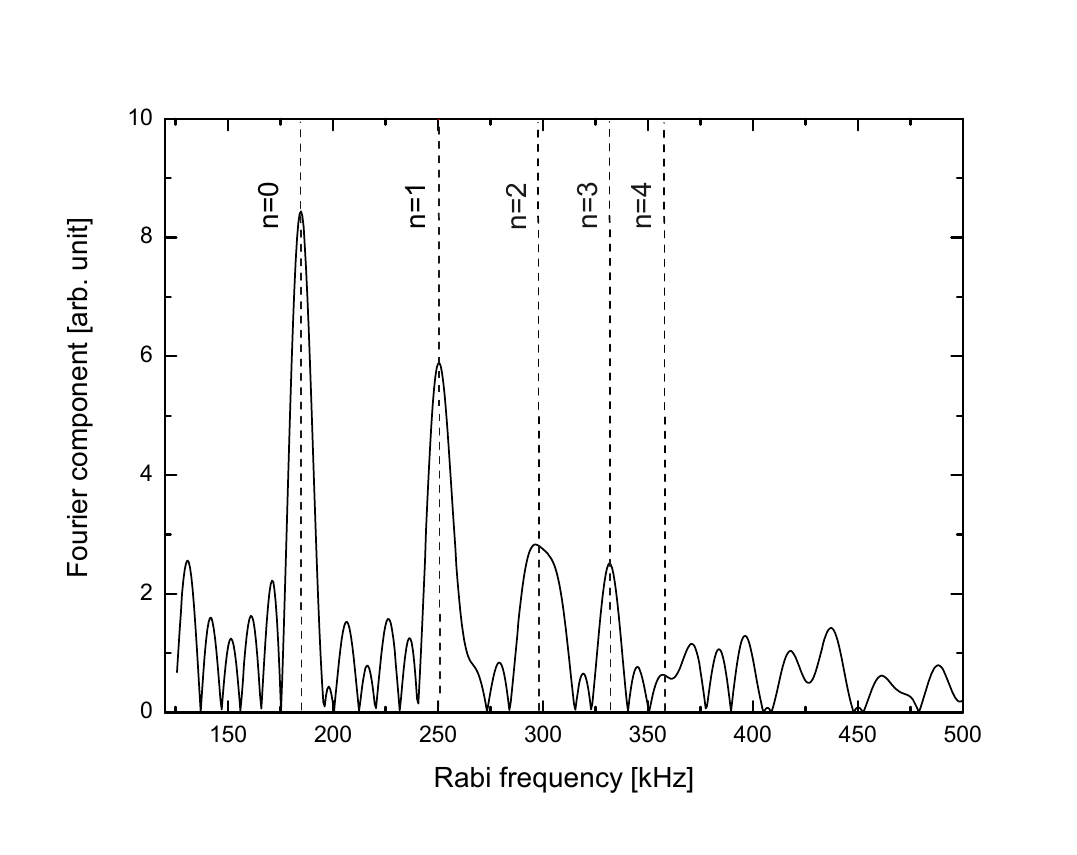} \caption{a)
Cosine transform of a R1/R2 pulse width scan on the BSB after a 3~ms delay between
cooling and probing. The dashed lines indicate the different flopping frequencies given
by the matrix element for the given transition.} \label{fftbsb}
\end{center}\end{figure}

A resulting spectrum is shown in Fig.~\ref{fftbsb}. Upon proper normalization, the peak
heights directly correspond to the occupation probabilities for the different phonon
numbers. This data can then be used to reconstruct the coherent dynamics, allowing for
the empirical inclusion of a coherence decay time~\cite{MEEKHOF1996}. This is done
according to
\begin{equation}
P_{\ket{\!\downarrow }}(t) =\sum_n\frac{P_n}{2}(A\;\cos{(\Omega_{n,n+1}\;t)}\;\e^{-\gamma\;t}+1),
\end{equation}
where $P_{\ket{\!\downarrow }}(t)$ is the probability for finding the ion in
$\ket{\!\downarrow }$, $P_n$ is the phonon number distribution, $\Omega_{n,n+1}$ is the
Rabi frequency pertaining the specific BSB transition, $A$ is the read-out contrast of
96\% and $\gamma$ is the coherence decay rate. The coherence time $1/\gamma$ is found
to be 280(20)~$\unitmicro{s}$. As Ramsey contrast measurements on the R1/CC transition
yielded a much longer coherence time, the additional decoherence either stems from
pulse area fluctuations or a reduced interferometric stability in the R1/R2 beam setup with respect
to the R1/CC geometry.
The phonon distribution is reconstructed for various waiting times after sideband
cooling in order to reveal the trap induced heating dynamics. The time dependent phonon
number distribution is shown in Fig.~\ref{phonons}, along with the resulting mean
phonon number. This directly gives the heating rate to be 0.3(1)
phonons/ms\footnote{This is about one order of magnitude better than earlier findings of
2.13 phonons/ms \cite{SCHULZ2008}, which is attributed to an improved trap voltage supply.}.

The corresponding Rabi oscillations on the carrier of the R1/R2 Raman transition are shown in
Fig.~\ref{coolstuff2}. Taking the phonon number after sideband cooling from the BSB Rabi
oscillations, we find excellent agreement with the measurements made on the carrier transition.

\begin{figure}[htp]\begin{center}
\includegraphics[width=1.0\textwidth]{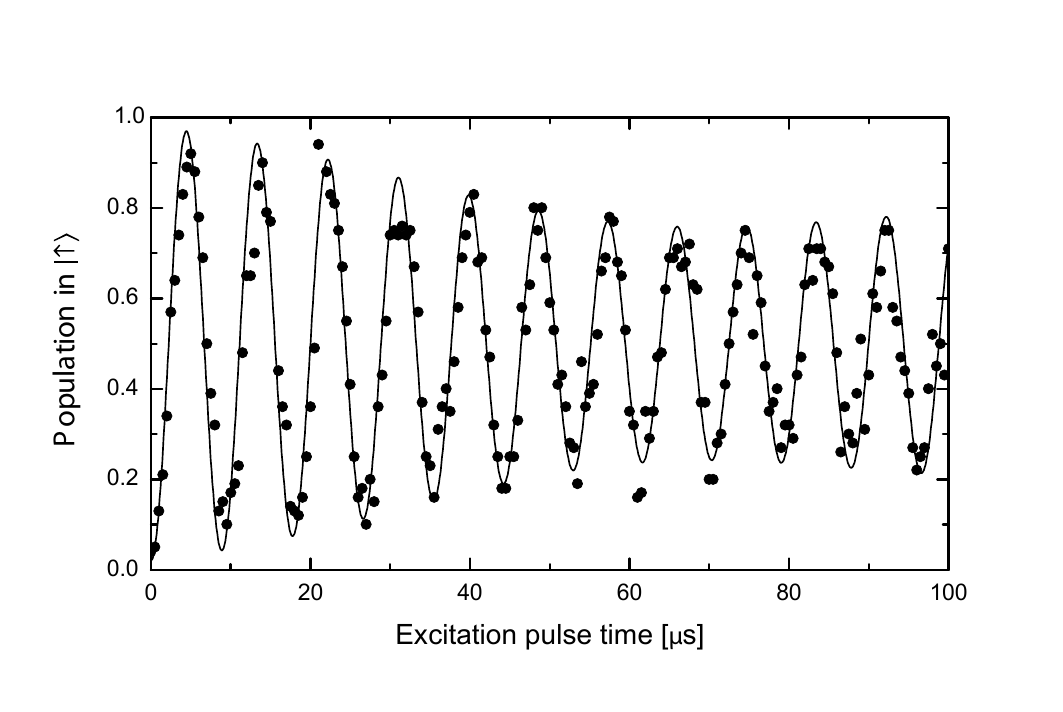}
\caption{Coherent dynamics on the R1/R2 carrier transition after sideband cooling. The
graph shows the population in the $\ket\downarrow$ versus pulse duration of a square
Raman excitation pulse directly after sideband cooling. The mean phonon number of
0.24 is used for fitting the data.} \label{coolstuff2}
\end{center}\end{figure}

\begin{figure}[htp]\begin{center}
\includegraphics[width=1.0\textwidth]{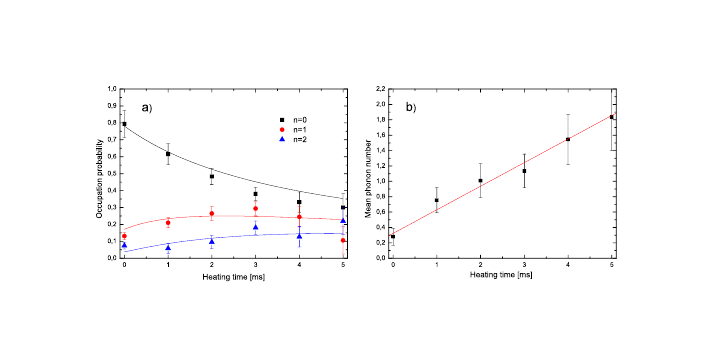}
\caption{a) Occupation probabilities $P_n$ for the lowest vibrational levels $n\leq 2$,
extracted from frequency spectra of the BSB pulse width scans (see Fig.~\ref{fftbsb})
after different waiting times. For comparison, the solid lines show the occupation
probabilities given by a thermal distribution $p(n)=\bar{n}^n/(\bar{n}+1)^{n+1}$, where
$\bar{n}(t)$ is given by a linear fit through the mean phonon numbers calculated from
the data. b) Mean occupation number $\bar{n}$ calculated from $P_n$ at different times
after cooling. The linear fit indicates a constant heating rate of
$\dot{n}=(0.3\pm0.1)/\textrm{ms}$. } \label{phonons}
\end{center}\end{figure}

\section{Outlook} \label{outlook}
In the future, the full control of a single spin qubit demonstrated here will be extended to two ion crystals. Then, a two-qubit quantum gate utilizing spin dependent light forces will be used for the
deterministic generation of Bell states.
Taking advantage of the multi-segmented micro ion trap we intend to split the entangled
two-ion crystal and investigate the separation of entangled states over distances of a
few mm. As the lifetime of entangled Bell states in the decoherence-free basis states
is long~\cite{HAEFFNER20052}, measured to be a few seconds in experiments using the Zeeman sublevels
of $^{40}$Ca$^+$, we expect to be able to generate many of these Bell states within their coherence
time. Protocols such as entanglement purification~\cite{JWPAN2003}, entanglement
swapping~\cite{RIEBE2008}, the generation of cluster states~\cite{BRIEGEL2001} will then be possible.

\vspace{1cm}

\textbf{Acknowledgements:}
We acknowledge financial support by the German science foundation DFG within the
SFB/TRR-21 and by the European commission within MICROTRAP (Contract No.~517675) and
EMALI (Contract No. MRTN-CT-2006-035369).

T%his can be circumvented by state selectively shelving population from one of the

\bibliographystyle{unsrt}
\bibliography{lit}

\end{document}